\documentclass[epj]{webofc}
\usepackage[varg]{txfonts}   
\usepackage{parskip} 
\begin{document}
%
\title{Multi-probe analysis of the galaxy cluster CL~J1226.9+3332:
hydrostatic mass and hydrostatic-to-lensing bias}
%
%

\author {\firstname{M.}~\lastname{Mu\~noz-Echeverr\'{\i}a}\inst{1}\fnsep\thanks{\email{miren.munoz@lpsc.in2p3.fr}\inst{\ref{LPSC}}}
   \and \firstname{R.}~\lastname{Adam} \inst{\ref{LLR}}
   \and  \firstname{P.}~\lastname{Ade} \inst{\ref{Cardiff}}
    \and  \firstname{H.}~\lastname{Ajeddig} \inst{\ref{CEA}}
   \and  \firstname{P.}~\lastname{Andr\'e} \inst{\ref{CEA}}
   \and \firstname{M.}~\lastname{Arnaud} \inst{\ref{CEA}}
   \and \firstname{E.}~\lastname{Artis} \inst{\ref{LPSC}}
   \and  \firstname{H.}~\lastname{Aussel} \inst{\ref{CEA}}
   \and  \firstname{I.}~\lastname{Bartalucci} \inst{\ref{Milano}}
   \and  \firstname{A.}~\lastname{Beelen} \inst{\ref{IAS}}
   \and  \firstname{A.}~\lastname{Beno\^it} \inst{\ref{Neel}}
   \and  \firstname{S.}~\lastname{Berta} \inst{\ref{IRAMF}}
   \and  \firstname{L.}~\lastname{Bing} \inst{\ref{LAM}}
   \and  \firstname{O.}~\lastname{Bourrion} \inst{\ref{LPSC}}
   \and  \firstname{M.}~\lastname{Calvo} \inst{\ref{Neel}}
   \and  \firstname{A.}~\lastname{Catalano} \inst{\ref{LPSC}}
   \and  \firstname{M.}~\lastname{De~Petris} \inst{\ref{Roma}}
   \and  \firstname{F.-X.}~\lastname{D\'esert} \inst{\ref{IPAG}}
   \and  \firstname{S.}~\lastname{Doyle} \inst{\ref{Cardiff}}
   \and  \firstname{E.~F.~C.}~\lastname{Driessen} \inst{\ref{IRAMF}}
   \and  \firstname{A.}~\lastname{Ferragamo} \inst{\ref{Roma}}
   \and  \firstname{A.}~\lastname{Gomez} \inst{\ref{CAB}}
   \and  \firstname{J.}~\lastname{Goupy} \inst{\ref{Neel}}
   \and  \firstname{F.}~\lastname{K\'eruzor\'e} \inst{\ref{LPSC}}
   \and  \firstname{C.}~\lastname{Kramer} \inst{\ref{IRAME}}
   \and  \firstname{B.}~\lastname{Ladjelate} \inst{\ref{IRAME}}
   \and  \firstname{G.}~\lastname{Lagache} \inst{\ref{LAM}}
   \and  \firstname{S.}~\lastname{Leclercq} \inst{\ref{IRAMF}}
   \and  \firstname{J.-F.}~\lastname{Lestrade} \inst{\ref{LERMA}}
   \and  \firstname{J.-F.}~\lastname{Mac\'ias-P\'erez} \inst{\ref{LPSC}}
   \and  \firstname{A.}~\lastname{Maury} \inst{\ref{CEA}}
   \and  \firstname{P.}~\lastname{Mauskopf}
\inst{\ref{Cardiff},\ref{Arizona}}
   \and \firstname{F.}~\lastname{Mayet} \inst{\ref{LPSC}}
   \and  \firstname{A.}~\lastname{Monfardini} \inst{\ref{Neel}}
   \and  \firstname{A.}~\lastname{Paliwal} \inst{\ref{Roma}}
   \and  \firstname{L.}~\lastname{Perotto} \inst{\ref{LPSC}}
   \and  \firstname{G.}~\lastname{Pisano} \inst{\ref{Cardiff}}
   \and  \firstname{E.}~\lastname{Pointecouteau} \inst{\ref{Toulouse}}
   \and  \firstname{N.}~\lastname{Ponthieu} \inst{\ref{IPAG}}
   \and  \firstname{G.~W.}~\lastname{Pratt} \inst{\ref{CEA}}
   \and  \firstname{V.}~\lastname{Rev\'eret} \inst{\ref{CEA}}
   \and  \firstname{A.~J.}~\lastname{Rigby} \inst{\ref{Cardiff}}
   \and  \firstname{A.}~\lastname{Ritacco} \inst{\ref{ENS},\ref{IAS}}
   \and  \firstname{C.}~\lastname{Romero} \inst{\ref{Pennsylvanie}}
   \and  \firstname{H.}~\lastname{Roussel} \inst{\ref{IAP}}
   \and  \firstname{F.}~\lastname{Ruppin} \inst{\ref{MIT}}
   \and  \firstname{K.}~\lastname{Schuster} \inst{\ref{IRAMF}}
   \and  \firstname{S.}~\lastname{Shu} \inst{\ref{Caltech}}
   \and  \firstname{A.}~\lastname{Sievers} \inst{\ref{IRAME}}
   \and  \firstname{C.}~\lastname{Tucker} \inst{\ref{Cardiff}}
   \and  \firstname{G.}~\lastname{Yepes} \inst{\ref{Madrid}}
}

   \institute{
     Univ. Grenoble Alpes, CNRS, LPSC-IN2P3, 53, avenue
des Martyrs, 38000 Grenoble, France
     \label{LPSC}
     \and
     LLR, CNRS, École Polytechnique, Institut Polytechnique de Paris,
Palaiseau, France
     \label{LLR}
     \and
     School of Physics and Astronomy, Cardiff University, Queen’s
Buildings, The Parade, Cardiff, CF24 3AA, UK
     \label{Cardiff}
     \and
     AIM, CEA, CNRS, Universit\'e Paris-Saclay, Universit\'e Paris
Diderot, Sorbonne Paris Cit\'e, 91191 Gif-sur-Yvette, France
     \label{CEA}
     \and
     INAF, IASF-Milano, Via A. Corti 12, 20133 Milano, Italy
     \label{Milano}
     \and
     Institut d'Astrophysique Spatiale (IAS), CNRS, Universit\'e Paris
Sud, Orsay, France
     \label{IAS}
     \and
     Institut N\'eel, CNRS, Universit\'e Grenoble Alpes, France
     \label{Neel}
     \and
     Institut de RadioAstronomie Millim\'etrique (IRAM), Grenoble, France
     \label{IRAMF}
     \and
     Aix Marseille Univ, CNRS, CNES, LAM, Marseille, France
     \label{LAM}
     \and
     Dipartimento di Fisica, Sapienza Universit\`a di Roma, Piazzale
Aldo Moro 5, I-00185 Roma, Italy
     \label{Roma}
     \and
     Univ. Grenoble Alpes, CNRS, IPAG, 38000 Grenoble, France
     \label{IPAG}
     \and
     Centro de Astrobiolog\'ia (CSIC-INTA), Torrej\'on de Ardoz, 28850
Madrid, Spain
     \label{CAB}
     \and
     Instituto de Radioastronom\'ia Milim\'etrica (IRAM), Granada, Spain
     \label{IRAME}
     \and
     LERMA, Observatoire de Paris, PSL Research University, CNRS,
Sorbonne Universit\'e, UPMC, 75014 Paris, France
     \label{LERMA}
     \and
     Univ. de Toulouse, UPS-OMP, CNRS, IRAP, 31028 Toulouse, France
     \label{Toulouse}
     \and
     Laboratoire de Physique de l’\'Ecole Normale Sup\'erieure, ENS, PSL
Research University, CNRS, Sorbonne Universit\'e, Universit\'e de Paris,
75005 Paris, France
     \label{ENS}
     \and
     Department of Physics and Astronomy, University of Pennsylvania,
209 South 33rd Street, Philadelphia, PA, 19104, USA
     \label{Pennsylvanie}
     \and
     Institut d'Astrophysique de Paris, CNRS (UMR7095), 98 bis boulevard
Arago, 75014 Paris, France
     \label{IAP}
     \and
     Kavli Institute for Astrophysics and Space Research, Massachusetts
Institute of Technology, Cambridge, MA 02139, USA
     \label{MIT}
     \and
     School of Earth and Space Exploration and Department of Physics,
Arizona State University, Tempe, AZ 85287, USA
     \label{Arizona}
     \and
     Caltech, Pasadena, CA 91125, USA
     \label{Caltech}
     \and
     Departamento de F\'isica Te\'orica and CIAFF, Facultad de Ciencias,
Modulo 8, Universidad Aut\'anoma de Madrid, 28049 Madrid, Spain
     \label{Madrid}
   }
\abstract{%
  We present a multi-probe analysis of the well-known galaxy cluster  CL~J1226.9+3332 as a proof of concept for multi-wavelength studies within the framework of the NIKA2 Sunyaev-Zel'dovich Large Program (LPSZ). CL~J1226.9+3332 is a massive and high redshift (z = 0.888) cluster that has already been observed at several wavelengths. A joint analysis of the thermal SZ (tSZ) effect at millimeter wavelength with the NIKA2 camera and in X-ray with the \textit{XMM-Newton} satellite permits the reconstruction of the cluster’s thermodynamical properties and mass assuming hydrostatic equilibrium. We test the robustness of our mass estimates against different definitions of the data analysis transfer function. Using convergence maps reconstructed from the data of the CLASH program we obtain estimates of the lensing mass, which we compare to the estimated hydrostatic mass. This allows us to measure the hydrostatic-to-lensing mass bias and the associated systematic effects related to the NIKA2 measurement. We obtain $M_{500}^{\rm{HSE}} = (7.65 \pm 1.03) \times 10^{14} M_{\odot}$ and $M_{500}^{\rm{lens}} = (7.35 \pm 0.65) \times 10^{14} M_{\odot}$, which implies a HSE-to-lensing bias consistent with 0 within 20$\%$.}
\maketitle
\section{Introduction}
\label{intro}
Galaxy clusters are excellent probes for cosmology, in particular for the understanding of large-scale structure formation processes \cite{boquet,planck1}. For this purpose, a precise knowledge of cluster masses is needed. The NIKA2 SZ Large Program (LPSZ) \cite{LPSZ} consists in 300 hours of Guaranteed Time dedicated to the observation and detailed analysis of 45 high redshift (z in 0.5-0.9) galaxy clusters taking advantage of the capabilites of the NIKA2 camera: a large field of view combined with high angular resolution at 150 and 260 GHz \cite{adam1,calvo,NIKA2-electronics,perotto}. These characteristics allow us to accurately map galaxy clusters via the thermal Sunyaev-Zel'dovich effect \cite{sz} and, together with high quality X-ray observations from \textit{XMM-Newton}, will enable the investigation of the SZ–mass scaling relation at high redshift for cosmological exploitation.

After a first analysis of a LPSZ cluster for a science verification study \cite{ruppin1} and a second one for proving the quality of NIKA2 in the most challenging case \cite{keruzore}, we present here a study on the robustness of the hydrostatic mass reconstruction combining SZ and X-ray data. We test the impact of systematic effects from the estimation of the filtering induced by the data processing pipeline and from the pressure model fitting. Systematic effects that may arise from other effects (e.g. density estimates at large radii) are not studied in this work.

\section{The galaxy cluster CL~J1226.9+3332}\label{sec-1}
The object of this analysis is the well-known galaxy cluster CL~J1226.9+3332, also known as PSZ2-G160.83+81.66. It is located at (R.A., Dec.)$_{J2000}$ = (12h26m58.37s, +33d32m47.4s) according to the X-ray peak from \cite{cavagnolo} with mass $M_{500}^{\rm{HSE}} \sim 5.70 \times 10^{14} M_{\odot}$ and redshift 0.888  \cite{psz2}, which makes it the highest-redshift cluster of the LPSZ sample. Discovered by ROSAT \cite{ebeling}, it has already been observed at several wavelengths \cite{adam2, maughan1,maughan2,jeetyson,joy,bonamente,allen,postman,zitrin1,romero}, appearing, from the first SZ observations with BIMA \cite{joy} as a very spherical cluster.


X-ray analyses using \textit{XMM-Newton} and Chandra data \cite{maughan1} found that the temperature of the cluster in a region at $\sim$ 40" to the south-west of the X-ray peak is much higher than the average temperature of the intra-cluster medium (ICM). Posterior SZ analyses with MUSTANG  \cite{korngut} and NIKA \cite{adam2,adam3}, as well as lensing \cite{zitrin1} and galaxy distribution studies \cite{jeetyson}, showed also the presence of such a substructure.
Various explanations have been proposed  \cite{jeetyson, maughan2}, suggesting mainly that the cluster consists of a two-halo system, which is being observed after the less massive cluster passed through the central one and the gas followed it. In short, CL~J1226.9+3332 shows a relaxed morphology at large scales and evidence of disturbance in the core.
\section{Map-making and data processing filtering}
\begin{figure}[h]
  \begin{minipage}[b]{0.45\textwidth}
    \includegraphics[trim={0.5cm 1.3cm 0cm 2.8cm},clip,scale=0.19]{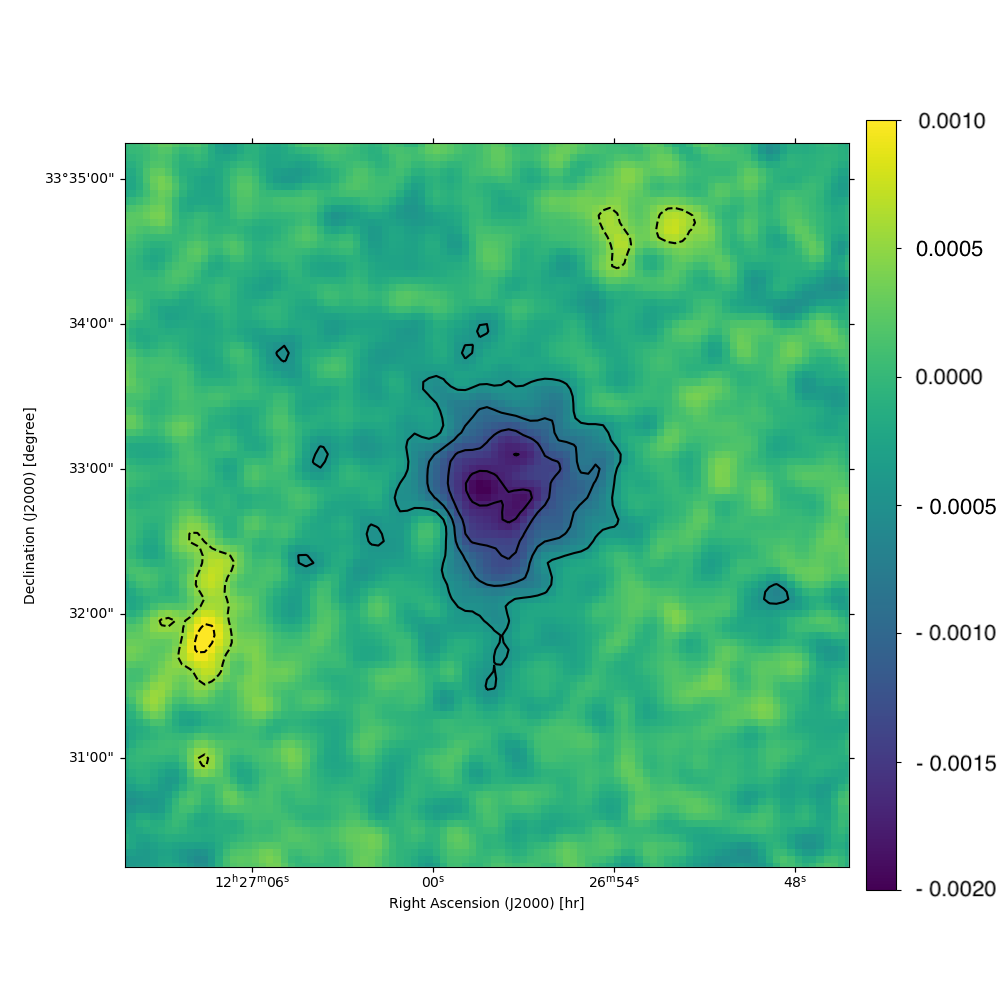}
  \end{minipage}
  \hfill
  \begin{minipage}[b]{0.45\textwidth}
    \includegraphics[trim={0.5cm 1.3cm 0cm 2.8cm},clip,scale=0.19]{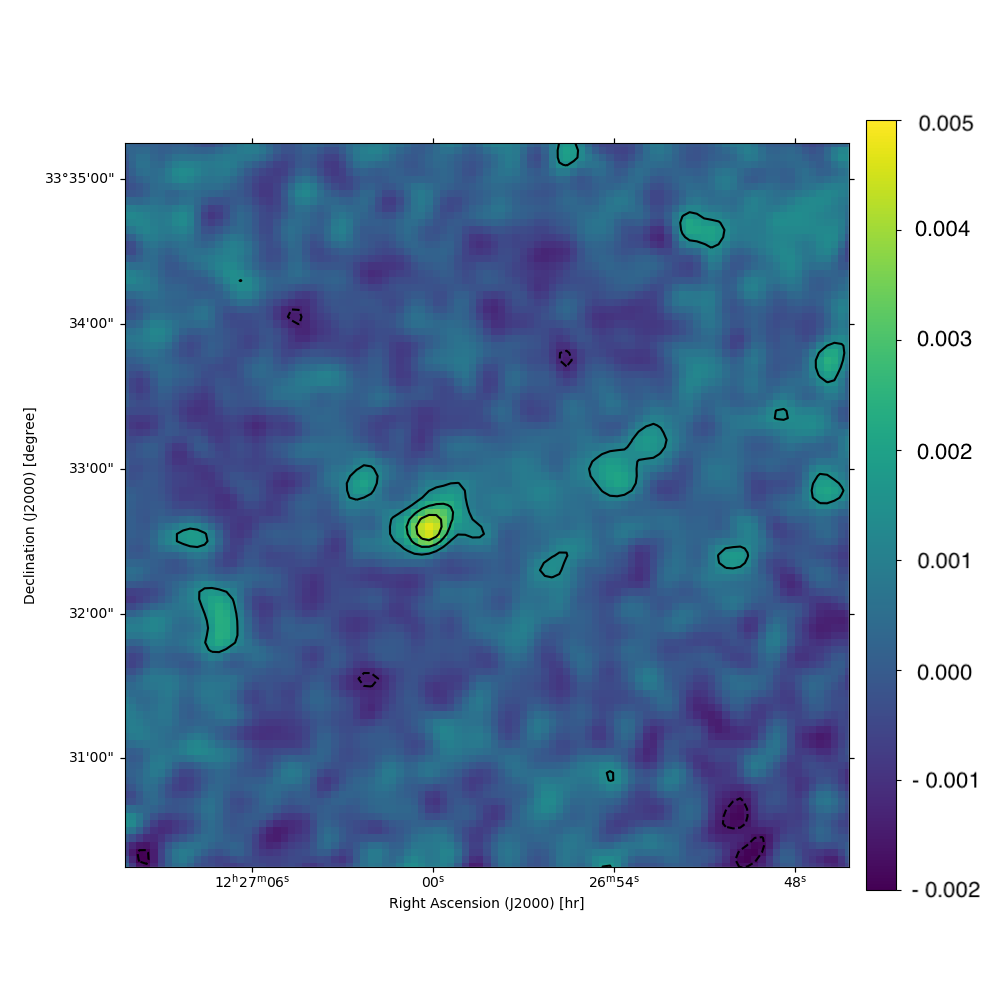}
  \end{minipage}
\caption{NIKA2 maps obtained with the LPSZ pipeline at 150 GHz (left) and 260 GHz (right) in Jy/beam units. Contours show S/N levels multiples of $\pm 3\sigma$. Both maps have been smoothed with a 10” FWHM Gaussian kernel.}  
\label{maps}
\end{figure}

The cluster was observed for 3.6 hours during the 15th NIKA2 science-purpose observation campaign (13-20 February 2018) as part of the NIKA2 Guaranteed Time at the IRAM 30-m telescope. The data were calibrated and reduced using the standard NIKA2 collaboration LPSZ pipeline described in \cite{perotto, keruzore,rigby}. We present in Figure~\ref{maps} the obtained maps at 150 and 260 GHz. The map at 150 GHz shows the cluster as a decrement of flux, contaminated by the positive signal from point sources. These point sources are also detected in the 260 GHz map. 

\begin{figure}[h]
  \centering
  \begin{minipage}[b]{0.450\textwidth}
    \includegraphics[trim={1.5cm 0cm 0.8cm 2cm},clip,scale=0.23]{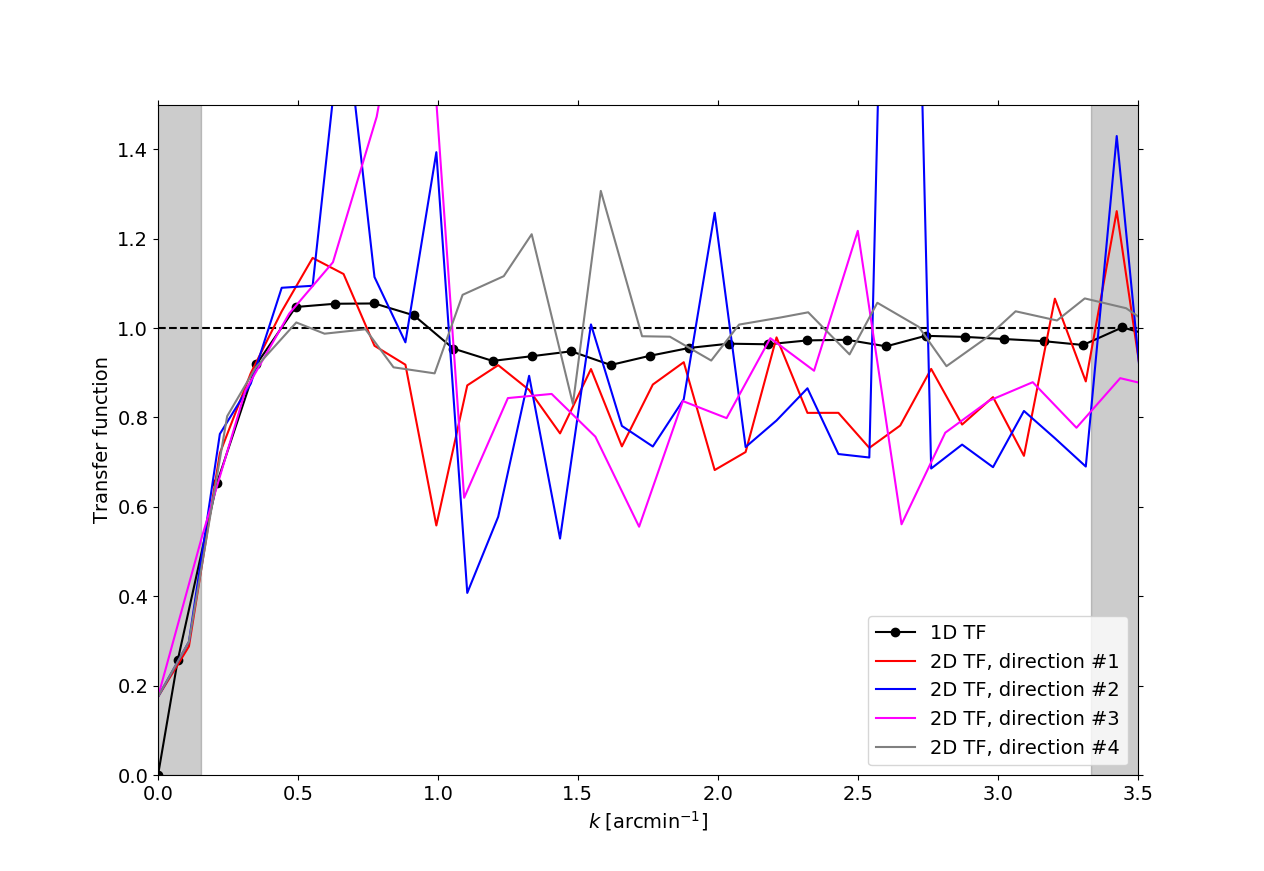}
  \end{minipage}
  \hfill
  \begin{minipage}[b]{0.45\textwidth}
    \includegraphics[trim={0.5cm 0cm 0.8cm 2cm},clip,scale=0.23]{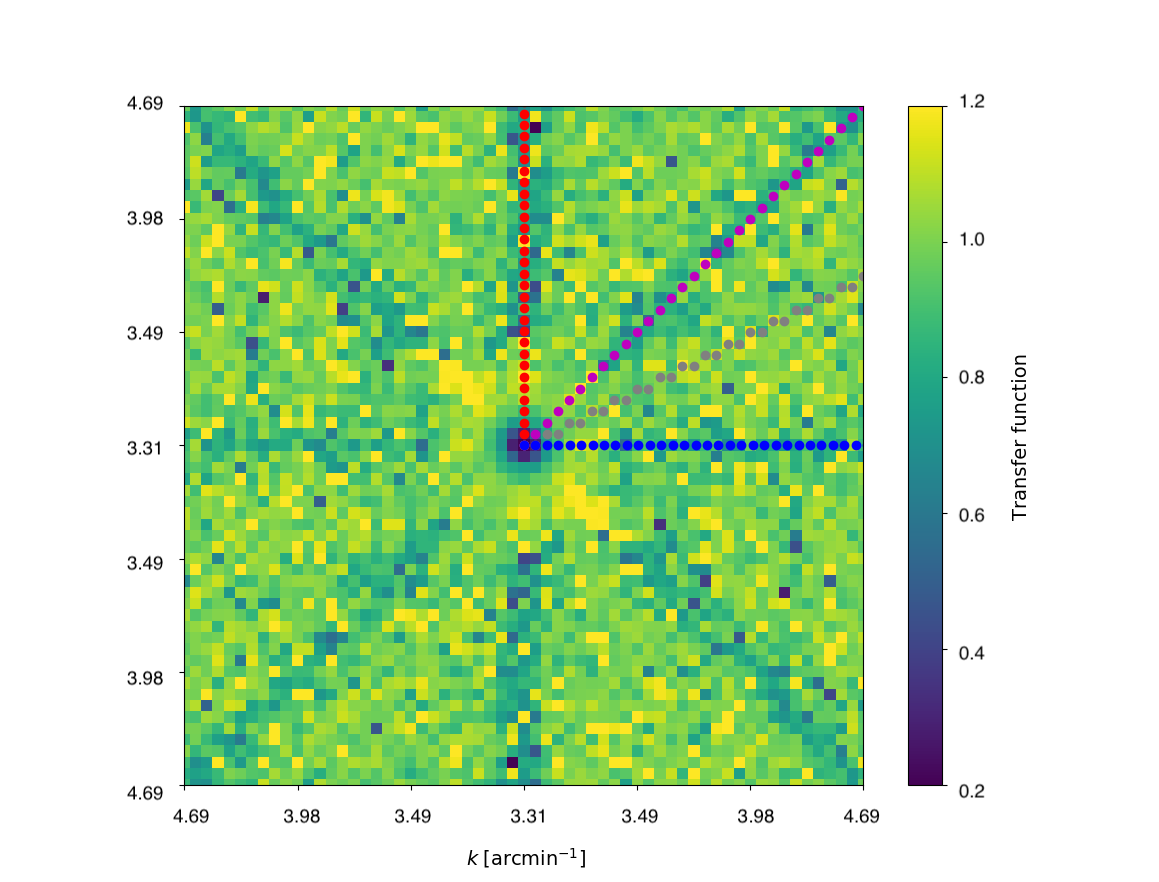}
  \end{minipage}
\caption{1D (left) and 2D (right) transfer functions representing the data reduction-induced filtering for the 150 GHz map in Figure~\ref{maps}. Color lines in the left panel show the value of the transfer function for different directions in the 2D TF, represented with the same colors in the right panel.}  
\label{tf}
\end{figure}

The filtering induced by the data processing in the 150 GHz map is computed using simulations via a transfer function (TF) that relates, in the Fourier space, the real astrophysical signal (the cluster) with the filtered one. The standard LPSZ analysis \cite{ruppin1,keruzore} uses one-dimensional transfer functions which are obtained by averaging the filtering in Fourier-domain annuli at a fixed angular scale, supposing therefore isotropy. The black line in the left panel in Figure~\ref{tf} shows the 1D TF for the 150 GHz map in Figure~\ref{maps}. In this work we test a new approach: a two-dimensional transfer function (2D TF) presented in Figure~\ref{tf}. That allows us to deal with anisotropy effects coming for example from the effect of scanning angles and the deviation from sphericity of clusters. Color lines in the left panel in Figure~\ref{tf} show the one-dimensional TFs for the different directions represented in the 2D TF and illustrate the impact of anisotropies in the angular frequency space.  

\section{SZ analysis: pressure profile determination}
The CL~J1226.9+3332 surface brightness map is proportional to the integrated electron pressure along the line of sight. To recover the latter, we perform a Monte Carlo Markov chain (MCMC) fit of a model to the 150 GHz map considering the Compton parameter signal associated with the cluster, the point sources and a zero level. The contamination from point sources at 150 GHz is estimated to account for their positive flux that may compensate the tSZ signal of the cluster. The details are given in \cite{keruzore, keruzore2}. In this work we performed the fit by adjusting two different pressure profile models: 1) a generalized Navarro-Frenk-White (gNFW, \cite{nagai}) and 2) a radially binned model. For the first one we fit for the 5 parameters of the model. The radially binned profile is defined by $P_{e}(r) = P_{i} \left( \frac{r}{r_{i}} \right) ^{-\alpha_{i}}$, where for each radial bin $r_{i}$ a value of the pressure $P_{i}$ and the slope $\alpha_{i}$ are fitted. Moreover, we repeated the fit considering either 1D or 2D transfer functions. In the following we take as reference the radially binned pressure model with the 1D transfer function.
\section{Hydrostatic mass and robustness tests}
Combining the spherical pressure profile resulting from the fit with the electron density profile obtained from X-ray data, we can compute, under the hydrostatic equilibrium (HSE) hypothesis \cite{ruppin1,keruzore, adam2}, the hydrostatic mass profile for CL~J1226.9+3332 and with it the probability distribution for $R_{500}^{\rm{HSE,}}$\footnote{Radius at which the mean mass (HSE mass in this case) density of the cluster is 500 times the critical density of the universe at its redshift.}. We note here that before deriving the pressure, we fitted a gNFW model to the radially binned profiles. This allowed us to have smooth pressure profiles to calculate the derivative of the profile for the mass estimate, but still testing the use of a different model when reconstructing the pressure from the maps.

\begin{figure*}[h]
   \begin{minipage}[b]{0.47\textwidth}
        \includegraphics[scale=0.27]{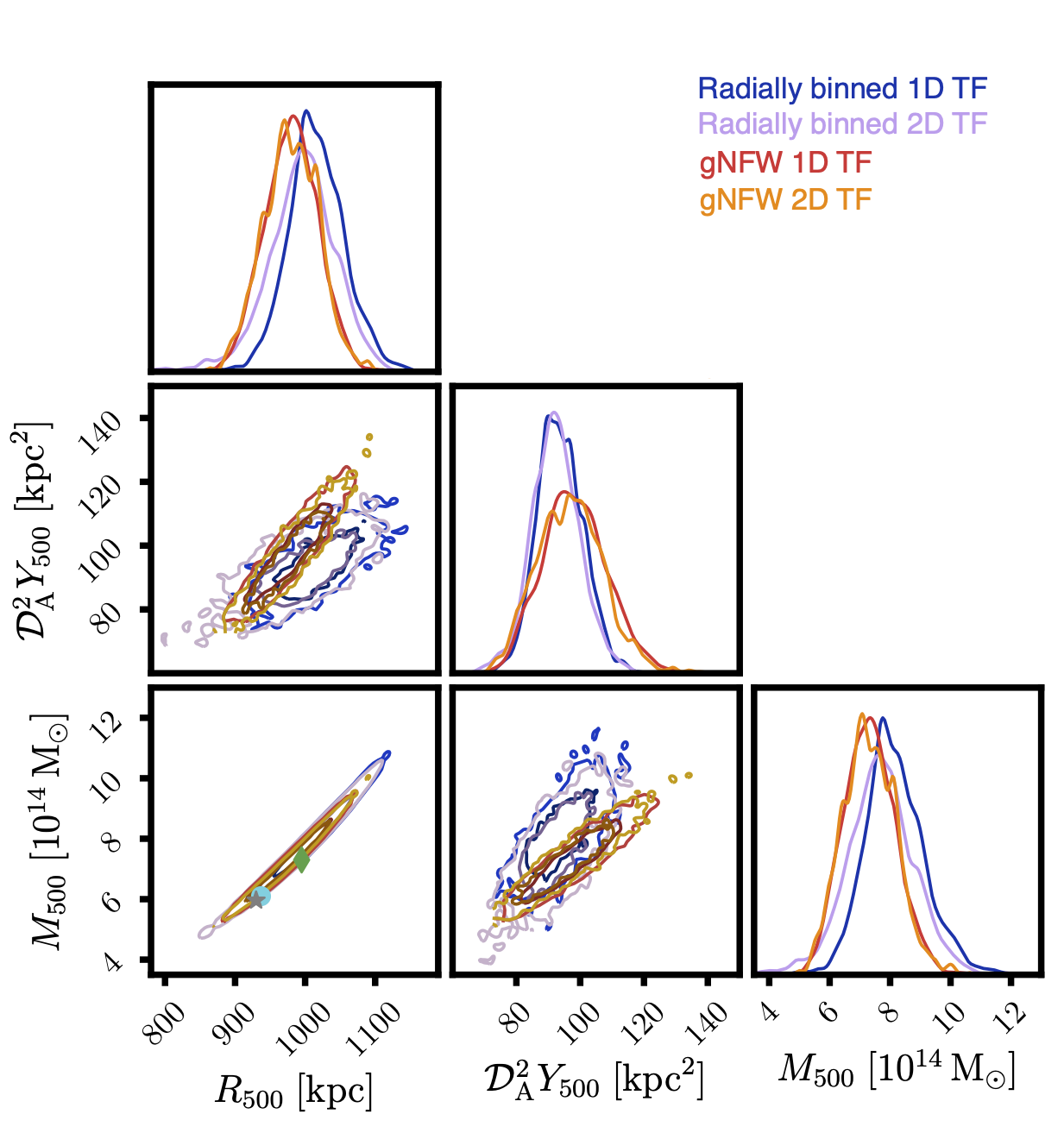}
    \end{minipage}
    \hfill
   \begin{minipage}[b]{0.47\textwidth}
   \includegraphics[scale=0.27]{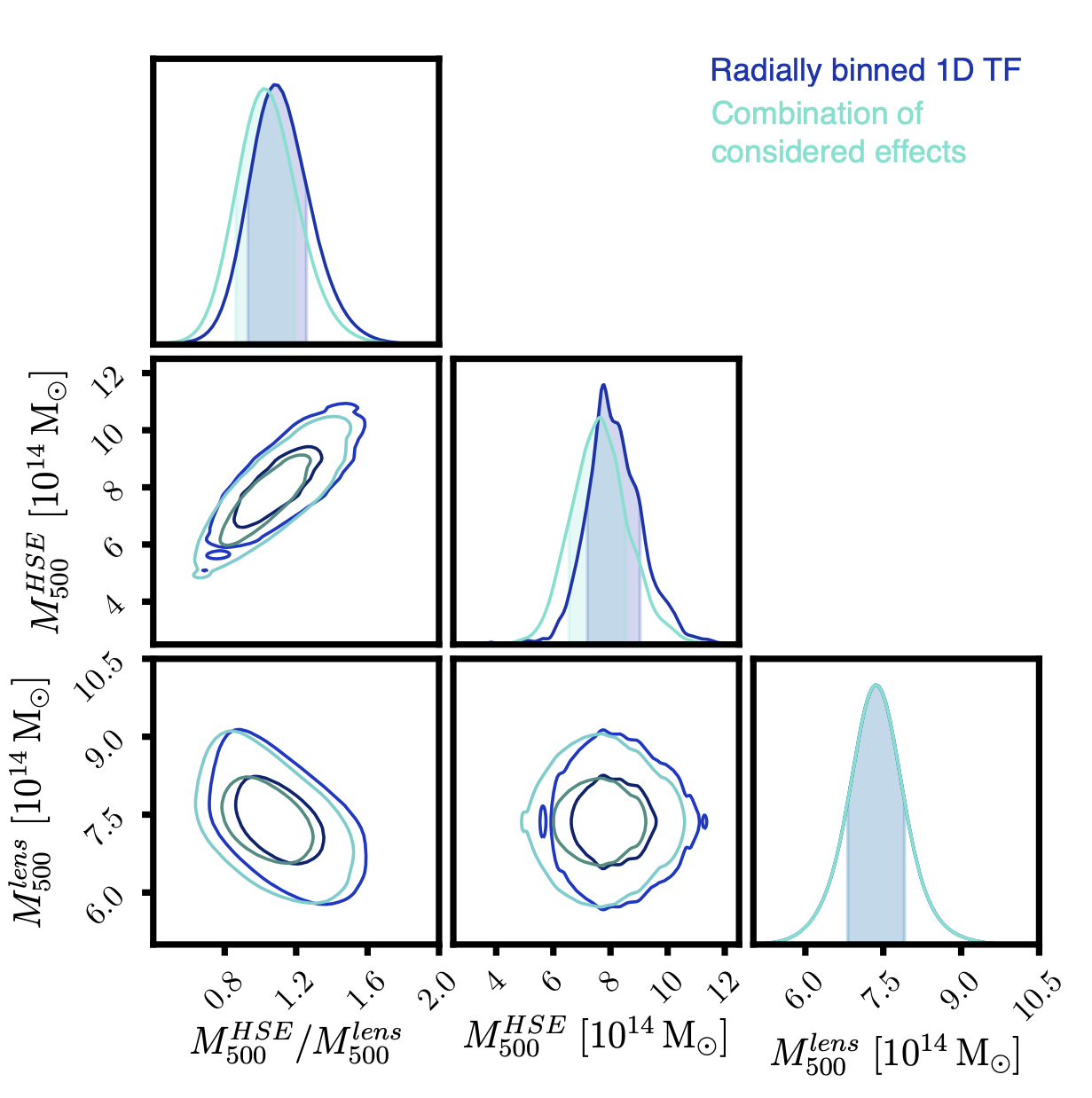}
    \end{minipage}
\caption{\textit{Left:} $R_{500}^{\rm{HSE}}$, $Y_{500}$ and $M_{500}^{\rm{HSE}}$ probability distributions for the radially binned pressure model with 1D TF in blue (reference), radially binned 2D TF in purple, gNFW 1D TF in red and gNFW 2D TF in orange. The grey star, cyan circle and green diamond correspond to the results from \cite{adam2} for the tested PPC, FPC and NNN models, respectively. \textit{Right:} Hydrostatic mass with respect to lensing mass at $R_{500}$. In dark shades of blue, 1$\sigma$ and 2$\sigma$ contours for $M_{500}^{\rm{HSE}}$ from the reference model  and in light shades of blue $M_{500}^{\rm{HSE}}$ as the combination of the four probability distributions in the left panel.}
\label{integrated}
\end{figure*}

We show in Figure~\ref{integrated} (left) the obtained probability distributions for $R_{500}^{\rm{HSE}}$, as well as the tSZ flux $Y_{500}$ and $M_{500}^{\rm{HSE}}$. These results are compared to previous HSE mass estimates obtained from NIKA maps \cite{adam2}. Our results give higher $R_{500}^{\rm{HSE}}$ and $M_{500}^{\rm{HSE}}$ values than expected from previous studies. Nevertheless, we observe a correlation between $R_{500}^{\rm{HSE}}$ and $M_{500}^{\rm{HSE}}$. This is expected since $R_{500}^{\rm{HSE}}$ and $M_{500}^{\rm{HSE}}$ are simultaneously obtained from the over-density definition giving directly proportional $(R_{500}^{\rm{HSE}})^{3}$ and $M_{500}^{\rm{HSE}}$.
Therefore, the comparison of probability distributions seems more representative when comparing different hydrostatic mass estimates. In the left panel in Figure~\ref{integrated} we also compare the results of the robustness tests we have performed starting from the 150 GHz map in Figure~\ref{maps}. We compare the probability distributions obtained with the reference model to the results using 2D transfer function instead. In the latter case, slightly lower values of $R_{500}^{\rm{HSE}}$ and $M_{500}^{\rm{HSE}}$ are favored. We also show results for the two pressure profile models. We find tighter constraints for the gNFW model when it is directly fitted to the map than when the radially binned profile is fitted and then gNFWs are adjusted to the resulting radially binned profiles. However, computing the pressure derivative in a consistent way (here always deriving from a gNFW), our mass estimate is robust. From the combination of the posterior probability distributions obtained for the four cases that we have considered, we measure an hydrostatic mass for CL~J1226.9+3332 of $M_{500}^{\rm{HSE}} = (7.65 \pm 1.03) \times 10^{14} M_{\odot}$, where $\sim$ 5-10 $\%$ of the uncertainty comes from the contribution of the tested systematic effects. 

\section{Hydrostatic-to-lensing mass bias}
We have performed an analysis of the convergence maps reconstructed from the CLASH data \cite{zitrin1} to obtain lensing mass estimates for CL~J1226.9+3332. Combining them we obtain $M_{500}^{\rm{lens}} = (7.35 \pm 0.65) \times 10^{14} M_{\odot}$. Details on this analysis will be given in \cite{munoz}. The hydrostatic-to-lensing mass bias $b_{\rm{HSE/lens}} = 1 - M_{500}^{\rm{HSE}}/M_{500}^{\rm{lens}}$, can be an interesting indicator of effects coming from the non-thermal contributions to the pressure of the cluster and systematic effects in mass estimations \cite{lovisari,pratt}. Supposing that they are uncorrelated estimates, we have combined the probability distributions obtained for $M_{500}^{\rm{HSE}}$ and $M_{500}^{\rm{lens}}$ to get a distribution of the ratio, which can be translated into a probability distribution of the hydrostatic-to-lensing mass bias for CL~J1226.9+3332. In the right panel in Figure~\ref{integrated} we present the probability distributions for $M_{500}^{\rm{HSE}}/M_{500}^{\rm{lens}}$, $M_{500}^{\rm{HSE}}$ and $M_{500}^{\rm{lens}}$. We have computed the ratio using the $M_{500}^{\rm{HSE}}$ from the reference model, as well as using the combination of the four probability distributions derived for $M_{500}^{\rm{HSE}}$, therefore accounting for all the considered effects. The figure shows that for both cases the bias is consistent with 0 within $\sim$ 20$\%$. 
\section{Conclusions}
Combined tSZ and X-ray analysis allow us to estimate the hydrostatic mass of this high redshift galaxy cluster, CL~J1226.9+3332. We have demonstrated that our hydrostatic mass estimates are robust against the application of the 1D or 2D transfer function. The reconstruction of the pressure profile derivative is a key element for the determination of the hydrostatic mass profiles, and needs to be dealt with carefully. This analysis has also been an important step towards developing a standard pipeline to compare and combine hydrostatic and lensing mass estimates. For CL~J1226.9+3332 we find a hydrostatic-to-lensing mass bias consistent with 0. 
\section{Acknowledgements}
\scriptsize{We would like to thank the IRAM staff for their support during the campaigns. The NIKA2 dilution cryostat has been designed and built at the Institut N\'eel. In particular, we acknowledge the crucial contribution of the Cryogenics Group, and in particular Gregory Garde, Henri Rodenas, Jean Paul Leggeri, Philippe Camus. This work has been partially funded by the Foundation Nanoscience Grenoble and the LabEx FOCUS ANR-11-LABX-0013. This work is supported by the French National Research Agency under the contracts "MKIDS", "NIKA" and ANR-15-CE31-0017 and in the framework of the "Investissements d’avenir” program (ANR-15-IDEX-02). This work has benefited from the support of the European Research Council Advanced Grant ORISTARS under the European Union's Seventh Framework Programme (Grant Agreement no. 291294). F.R. acknowledges financial supports provided by NASA through SAO Award Number SV2-82023 issued by the Chandra X-Ray Observatory Center, which is operated by the Smithsonian Astrophysical Observatory for and on behalf of NASA under contract NAS8-03060.} 



%
%
%

\end{document}